# International news flows theory revisited through a space-time interaction model

## Application to a sample of 320000 international news stories published through RSS flows by 31 daily newspapers in 2015

## Abstract

This paper proposes a quantitative model of the circulation of foreign news based on a gravity-like model of spatial interaction disaggregated by time, media and countries of interest. The analysis of international RSS news stories published by 31 daily newspapers in 2015 demonstrates, first, that many of the laws of circulation of international news predicted half a century ago by Galtung and Ruge and by Östgaard are still valid. The salience of countries in media remains strongly determined by size effects (area, population), with prominent coverage of rich countries (GDP/capita) with elite status (permanent members of UNSC, the Holy See). The effect of geographical distance and a common language remains a major factor of media coverage in newsrooms. Contradicting the flat world hypothesis, global journalism remains an exception, and provincialism is the rule. The disaggregation of the model by media demonstrates that newspapers are not following exactly the same rules and are more or less sensitive to distance, a common language or elite status. The disaggregation of the model by week suggests that the rules governing foreign news can also be temporarily modified by exceptional events that eliminate the usual effects of salience and relatedness, producing short periods of "global consensus" that can benefit small, poor and remote countries. This paper concludes by recommending the use of a sample of carefully chosen diversified media rather than a large aggregation of data for global studies.

## Introduction

The research field of international news flow theory (INFT) was initiated 50 years ago by the joint publication of two papers in the same issue of the *Journal of Peace Research* (Galtung & Ruge, 1965; Östgaard, 1965). Both papers proposed general laws likely to explain the differences in media coverage of countries in international news and to underline the importance of the scientific and political challenges to be addressed by their followers.

In the wide set of research possibilities opened in 1965, the analysis of the global salience of countries in international news flows appears to be a specific field of cross-disciplinary research between specialists from communications, geography and political science. Indeed, the evaluation of the international salience of countries is measured through an aggregation of news over a period of time and over a set of media. In contrast to many analyses conducted on a specific subject (e.g., Ebola, earthquakes, civil wars, climate change), the evaluation of the salience of countries is necessarily quantitative and "blind" – at least in the first step of analysis – to the qualitative content of events. This abstraction is precisely the reason why this specific branch of media studies has attracted the interest of researchers from other disciplines working on quantitative measures of state power or on a spatial interaction model of migratory or economic flows.

It would be of major interest for specialists in global studies and peace research to combine in the same analysis different types of flows to propose new visions on the trends of globalization and regionalization in the contemporary world. In addition, international news flows undoubtedly have a major role to play in the analysis of international crisis and conflicts. However, can we actually transpose methods used for the analysis of migratory and trade flows to media flows? Are they predictable without introducing specific variables related to the specific events occurring during the period of observation ? Can we apply to media studies the spatially integrated approach of social sciences suggested by Goodchild & Janelle (2004) ?

To answer these questions, we first provide a brief state of the art concerning INFT and its relation with more general models of spatial interaction used by economists, geographers and political scientists. Then, we present our hypotheses and the data used for their verification. Subsequently, we formalize the model and discuss the results.

## The state of the art

The joint publication of two papers in the same issue of the *Journal of Peace Research* in 1965 (Galtung & Ruge, 1965; Östgaard, 1965) is considered the birth of the INFT paradigm. Indeed, these two papers realized a crystallization of previous studies through the elaboration of a bulk of hypothetical "factors" and "laws" that was supposed to govern the publication of foreign news. The verification of the laws proposed by these authors defined an exciting and ambitious research agenda at the crossroad of many disciplines, with high potential political consequences: Fifty years later, we can acknowledge that this prediction was partly realized. Indeed, an enormous number of publications have been produced by researchers in media studies, political science, international relations, sociology and even geography either validating (Kim & Barnett, 1996; Wu, 1998, 2000, 2003, 2007 ;  Shoemaker, 2006 ; Segev, 2010,2014,2017), or criticizing (Palmer, 2000 ; Harcup & O'Neill, 2001 ; Richardson, 2007 ; O'Neill & Harcup, 2009) what is commonly designated *Galtung's laws*.

## The 12 rules of Galtung and Ruge

In their seminal paper, Galtung and Ruge (1965) declare that they are not interested in the detailed steps of the transmission chain of events to the newspaper desk by means of journalists in the field, local newspapers, major press agencies, editors, etc. These factors are important but are treated *in abstracto* to focus only on the result and to formulate the general problem as follows: "How do events become news?"  In this regard, Galtung suggests to describe the news circulating around the world as a giant set of broadcasting stations:

Imagine that the world can be likened to an enormous set of broadcasting stations, each one emitting its signal or its program at its proper wavelength. The emission is continuous, corresponding to the truism that something is always happening to any person in the world. Even if he sleeps quietly, sleep is "happening" – what we choose to consider an "event" is culturally determined. The set of world events, then, is like the cacophony of sound one gets by scanning the dial of one's radio receiver, and particularly confusing if this is done quickly on the medium-wave or short-wave dials. Obviously this cacophony does not make sense, it may become meaningful only if one station is tuned in and listened to for some time before one switches on to the next one. Since we cannot register everything, we have to select, and the question is what will strike our attention. (Galtung & Ruge, 1965, p.65)

This metaphor is immediately followed by the enumeration of the first step of 8 proposals (F1-F8) that are not presented as a theory but as "some obvious implications of this metaphor". This first group of factors, according to the authors, is based on abstract laws of psychology that are supposed to be at work in all mankind. As noted by the authors, these eight factors "are held to be culture-free" and should not introduce variations between media located in the North or the South, in the East of the West, in the center of the periphery. A second group of factors (F9-F12) is therefore added to cover these variations related to culture, economics and politics. After having described each of them, the authors finally present a simplified table of the 12 factors (Table 1).

----------------------------

Add Table 1 here

----------------------------

These rules have been crucial for the development of media studies but – and it is our original perspective – they are also strongly connected with theoretical research developed in various domains of social sciences, in particular in the fields of economy, political sciences and geography. As we will discuss later (see. *Hypotheses*) the effects of size, similarity, distance and barriers on interaction does not only concerns media flows but has been also successfully applied to economic flows, trade flows, political relations …

## The additivity hypothesis

The crucial point that Galtung and Ruge prominently underline is the fact that these factors do not produce effects in an isolated manner and generally must be combined. From this perspective, a crucial assumption of additivity concerning the cumulative effect of the different factors is proposed: *The higher the total score of an event, the higher the probability that it will become news and even make headlines*. This hypothesis is crucial because it implies that one factor of newsworthiness can be balanced by another. For example, a remote and poor country such as Nepal has a low level of newsworthiness for media located in Europe or the USA because it is a country that is remote (F4) and poor (F9). However, these handicaps can be balanced by events such as the earthquake of 2015, which was especially intense (F1), unexpected (F6) and negative (F12). The followers of Galtung who have developed a model of salience of countries (Wu, 1998, 2000 ; Segev, 2010,2014)  have logically concluded from this rule of additivity that the factors of newsworthiness cannot be analyzed one by one. It is necessary to combine their effects in multiple regression model.

In general, however, only some rules have been intensively tested, with a particular focus on the effects of meaningfulness (F4), consonance (F5) elitism (F9, F10), personalization (F11) and negativity (F12). However, only very few studies have tried to address the complete set of rules, which means that they have particularly ignored (F1) the law concerning the hypothesis of a minimum threshold and the law (F7) concerning the persistence of the signal (see. Table 1). To the best of our knowledge, the only empirical verification of the additivity hypothesis on the full set of laws (F1-F12) was performed by Peterson based on a set of 5466 events of which only 15% were subject to publication as news in *The Times* of London. Indeed, these results revealed a positive correlation between the number of factors of newsworthiness verified by an event and the probability of publication (Peterson, 1981). One possible reason why it is difficult to consider the whole set of rules is the fact that deciding on the status of news in an objective manner according to the rules proposed by Galtung and Ruge is not always obvious. Having tried to apply the rules on a sample of news with double checking, some authors have considered that "a *number of Galtung and Ruge's factors appear to be problematic to identify while others may be identifiable but less in any intrinsic properties of a potential news story and more in the process of how a story has been constructed or written up*"(Harcup & O'Neill, 2001, p. 277).

## Östgaard's theory of the "free flows of news"

Östgaard's contribution (Östgaard, 1965)  should be rediscovered with regard to at least one particular point that is not present in Galtung's laws(Galtung & Ruge, 1965),: the introduction of an implicit econometric model of competing news that takes the form of a quasi-zero-sum game. According to Östgaard (1965) , the free flow of news is certainly hampered by political factors such as censorship, propaganda or subtler forms of "news management" by governments or international news agencies. However, he considers that economic factors also play a major role, particularly through the feedback created by audience on the media selection of news. In this regard, Östgaard suggests that they are no substantial differences between media located in capitalist or socialist countries:

We do maintain that for whatever reason the news is published, it will be intended to reach an audience. Whether the newsman is working for The Times or for Tass or for the Daily Yell, he will try to capture the attention of the reader, to interest him. (Östgaard, 1965, p. 52)

The theoretical framework proposed by Östgaard thus implies that the newsworthiness of an event should not necessarily be defined in absolute terms but rather in relative terms as a **choice model subject to constraints**. This idea of "choice" is clearly presented in the definition of the processing of news, which is described as a kind of global stock exchange:

Every minute of the day, thousands of men and women are at work, gathering, transmitting and presenting the news to the public. The decisions they make, when discarding most material available, when choosing what is to be presented, and when presenting in the way they consider best, will be called the processing of news. (Östgaard, 1965, p.40)

These competitions between news can explain what he called *news barriers*, i.e., the existence of continuity or discontinuity in the coverage of event stories. At every moment, events occurring in different countries are engaged in a form of competition for publication, and it is not only the newsworthiness of an event but also the existence of an international market of other competing events simultaneously occurring that matters. For example, it is very clear that the growing media interest in the deaths of migrants crossing the Mediterranean Sea between Libya and Italy at the end of April 2015 was strongly interrupted when the earthquake in Nepal occurred. Conversely, the September 2017 earthquake in Mexico failed to reach a peak of interest because of the pre-existing media focus on the disaster induced by Hurricane Jose.

# The aggregation of foreign news and measures of the salience of countries

Choosing between the event-oriented approach or the country-oriented approach has been a crucial dilemma for the followers of Galtung and Ruge and of Östgaard.

The event-oriented approach is based on a selection of foreign news related to a specific topic for which it is possible to define a finite and possibly objective list of events occurring in the "real" world. One of the most interesting areas of research from this perspective is the study of the mediatization of earthquakes, for which objective measures of magnitude or victims are regularly published. It is then possible to analyze the level of newsworthiness according to the different laws postulated by Galtung (Koopmans & Vliegenthart, 2010; Le Texier, Devès, Grasland, & De Chabalier, 2016). The first problem with this approach is the fact that it is impossible to establish an objective list of events related to the topic (e.g., social movements). However, the most important issue with the segmented approach is the fact that we cannot assume that the different related topics in international events belong to a separate arena (Nossek, 2004; Wanta & Hu, 1993).

The country-oriented proposes a verification of Galtung's laws through the aggregation of international news by country, disregarding the topic of the events responsible for the unequal salience of countries. Historically, many authors have contributed to this field of research (Kim & Barnett, 1996; Shoemaker, 2006), and empirical verifications on large samples of data, particularly by Haoming Denis Wu (1998, 2000, 2003, 2007) and Elad Segev (Segev, 2010, 2014, 2017), have been performed. This research has confirmed that the global salience of countries is strongly determined by country size, elite status and various factors of relatedness, including geographical, historical, economic or cultural distances. These authors have also introduced new factors that are not present in Galtung's laws such as the existence of flows between countries (e.g., trade, migration, tourism, cultural and knowledge exchange) and the presence or absence of news agencies. However, they have generally used predictive models without systemic constraints, which means that they have not considered Östgaard's hypothesis regarding competing news (Grasland, Lamarche-Perrin, Loveluck, & Pecout, 2016). One of the most important caveats revealed by empirical studies is the difficulty of identifying a so-called normal period, where the salience of countries will not be biased by "exceptional" events.

## Hypotheses

The originality of our proposal is to propose a theoretical model of circulation of international news where only structural factors are involved and where we deliberately ignore the explanatory variables related to a specific conjectural situation. Of course, we do not ignore that news from conflict ridden countries (such as Pakistan, Afghanistan, Libya, Iraq, Syria and so on) usually have prominence by virtue of regional and international political factors. But we want to evaluate precisely what can be explained by permanent factors (i.e. constant over a temporal period of minimum 20-years) and analyze precisely the residuals of these structural model for the identification of events that has attracted the attention of media.

The model that we propose is therefore voluntarily limited to the general parameters of size and distance classically used in spatial interaction models of the gravity type that are applied to the description of migratory flows and trade flows. These two factor families are typically the classical ingredients in spatial interaction models used by economists or geographers to model migratory flows or trade flows. They are inherited from a theoretical tradition known as the *gravity model* that was empirically formulated at the end of the nineteenth century by Ravenstein in a seminal work on *the laws of migration (Ravenstein, 1885; Tobler, 1995)*.

The first mathematical equation of the gravity model ($F_{ij} = k.P_iP_j/D_{ij}$) was derived from Newton's law and published by Reilly in the famous book titled *Human Behavior and the Principle of Least Effort* (Zipf, 1949). The gravity model was immediately applied by economists and geographers to migratory flows and trade flows. However, in Zipf's view, the model was more general. In addition, it is interesting to point to an earlier publication of Zipf in *The American Journal of Psychology* where he proposed an application of this model to the determinants of the circulation of information, with a specific focus on news published by The Chicago Tribune and The New York Times (Zipf, 1946). From Zipf's perspective, analyzing the circulation of information was crucial because he considered that movements of people or goods cannot occur without a previous circulation of information (Hagerstrand, 1966) and the location of intervening opportunities (Stouffer, 1940)

Further developments of Zipf's proposal have led to a strong improvement of the initial model, particularly not only with the introduction of discontinuities in the distance effect in the case of cross-border flows (Bröcker & Rohweder, 1990; Linneman, 1966; Mackay, 1958) but also with the addition of systemic constraints in the models, considering the competition effect of opportunities of relation (Dorigo & Tobler, 1983; Wilson, 1967). As a whole, the development of the gravity model realized in recent decades (Fotheringham & O'Kelly, 1989; Sen, Smith, & Nijkamp, 1995) offers a solid theoretical and methodological basis for the spatial analysis of social facts in general (Grasland, 2009) and the circulation of news flows in particular.

## The quantitative factors of salience (size effect)

We propose to first define the quantitative salience of countries $S_p$ as a combination of three complementary factors of size that increase the probability of the appearance of events of interest to media (eq. 1):

- **Geographical size**, as measured by the land area of the country (Source: World Bank), defines the first dimension of salience. We assume that the many events of interest to international news are related to the size of the territory, which is typically the case of so-called natural events (e.g., earthquakes, volcanos, storms), but it also remains true for economic or social phenomena.

- **Demographic size** is measured by the resident population of countries (Source: World Bank). As it is correlated to geographical size, we introduce population density into the model to evaluate the specific effect of the population located in a territory. Combined with the previous effect, the two parameters describe a utopic situation of equal coverage of mankind by media.

- **Economic size** is measured by GDP in purchasing power parity (Source: World Bank). It is introduced into the model as GDP per capita to evaluate the specific preference for most developed countries, all things being equal with the equal coverage of territories defined by the previous parameters.

$$S_p = \left(SUP_p\right)^{\alpha_1} \times \left(\frac{POP_p}{SUP_p}\right)^{\alpha_2} \times \left(\frac{GDP_p}{POP_p}\right)^{\alpha_2} \qquad \text{(eq. 1)}$$

## The qualitative factors of salience (elite status)

We assume that the three quantitative salience factors are not sufficient to cover the specific situation of some countries that are structurally more important than others in international news flows, even when no specific events occur within their borders. It is therefore necessary to introduce some qualitative parameters that define the elite status of specific countries in the international system $E_p$ (eq. 2):

- First, the **five permanent members (P5) of the United Nations Security Council (UNSC)** are likely to be more involved than other countries in international news because they are associated with the majority of decisions in cases of conflict, with veto power. The Syrian crisis and the nuclear agreement with Iran are two examples of events where the P5 have benefited from a specific level of mediatization.
- The **other members of the G20 (G14)** (i.e., the 14 countries excluding the P5 and the EU) have been introduced into the model to verify whether they can also be considered elite countries when we exclude the specific case of the P5.
- The **Vatican effect** has been introduced as an ad hoc parameter considering the exceptional positive residual that appeared for this country in the first simulation of our model, especially in newspapers from Latin America. It is fair to admit that the introduction of such and *ad hoc* explanatory variables contradict to some extent our objective to introduce only structural variables. But contrary to war and conflict (such as Afghanistan, Iraq, Syria, Lebanon, Pakistan, etc), the effect of Holy See appears has something much more structural and permanent.

$$E_p = (\beta_1)^{PM5_p} \times (\beta_2)^{G14_p} (\beta_3)^{VAT_p} \qquad \text{(eq. 2)}$$

## The relatedness factors (quantitative or qualitative)

The relatedness function describes the factors likely to increase or reduce the probability that the international desk of media outlet (m) will select news about country (p), all things being equal with the salience effect. After various tests including other factors of proximity (a common border, belonging to the same continent, a former colonial relation), we have decided to limit the model to the two factors that are the most predictive and the least correlated (eq. 3):

- **Geographical proximity (DIS)** is measured by the grand circle distance between the capital of the host and guest countries (Source: CEPII). We use inverse transformation (1/Distance in km) to obtain a factor of proximity.
- **Linguistic proximity (LAN)** is measured by the fact that the host and guest countries share a common official language or a common language spoken in each country by 20% of the population (Source: CEPII). This effect of a common language can be explained by the facility of communication. However, it can also be considered a general proxy of various cultural and historical factors of proximity.

$$R_{mp} = \left(\frac{1}{DIST_{mp}}\right)^{\gamma_1} \times (\gamma_2)^{LANG_{mp}} \qquad \text{(eq. 3)}$$

## The kick-off effect

Finally, we introduce a parameter of time dependence to measure whether media have published a minimum number of news stories about the country of interest during the previous period of time. This parameter is clearly a reference not only to the 7th rule of Galtung ("If one signal has been tuned in to, the more likely it will continue to be tuned in to as worth listening to") but also to the idea of news barriers formulated by Östgaard. Here, we use a dummy variable (TIM) that measure a "kick-off" effect for countries still active in media (eq. 4).

$$T_{mpt} = (\delta)^{TIM_{mpt-1}} \qquad \text{(eq. 4)}$$

## The general form of the model

We add to the model an offset variable that indicates the number of news stories sent by each media outlet during each week if the number is not equal (eq. 5):

$$F_{mpt} = F_{mt} \times S_p \times E_p \times R_{mp} \times T_{mpt} \qquad \text{(eq. 5)}$$

The parameters obtained by the model can therefore be used to estimate a choice model that indicates the average probability for each country to be chosen by each media outlet during each week (eq. 6):

$$prob(mtp|mt) = \left(\frac{F_{mtp}}{F_{mt}}\right) = S_p \times E_p \times R_{mp} \times T_{mpt} \qquad \text{(eq. 6)}$$

## The exclusion of flow-based and event-based factors

In contrast to the majority of media studies specialists, we have voluntarily excluded two groups of explanatory variables.

- **The factor of relatedness based on flows**: Many studies produced by specialists in media flows introduce a contemporary interaction in models such as bilateral trade flows, the number of foreigners present in host countries coming from guest countries or, conversely, the number of tourists from host countries visiting guest countries (Wu, 1998, 2000 ; Segev, 2016). These factors are generally characterized by high explanatory power and are always very significant from a statistical perspective . However, is it logically consistent to introduce them into the model? Trade, tourism and migratory flows are themselves strongly related to size and distance factors, which means that a statistical bias will be introduced in the model. And what is the most important, migration, trade or tourism cannot be considered as independent factor explaining news flows because they are themselves depending from the information field linking countries (Hägerstrand, 1966). As demonstrated by Snyder & Kick (1975), the world system is organized by a multiple-network analysis of transnational interactions. And it is not possible to isolate one type of linkage for the explanation of another one without introducing very complex feed-back effects.

- **The factors of salience related to the number of "events" occurring in a country during the period**. As explained in the beginning of these section we have voluntary excluded these factor because we want to build a model where only structural factors are involved and where events are analyzed through the residuals. But another important reason of exclusion of these event factors is the difficulty to find objective measure of the presence of war, natural disasters, economic crises in countries, especially when we try to disaggregate the model by short period of time as we do here on a weekly basis.

## Data

A detailed presentation of the data used in these paper is available in the *Supplementary Material*, and more details can be found in Grasland & al. (2016). Here, we therefore present only a short summary of the data collection and harmonization methodology.

## Presentation of the sample

The aim of the data collection is to build a 3-D cube describing the circulation of international information within a set of media (m) with regard to a set of places (p) during several periods of time (t).

In this paper, we use a sample of 321269 news stories published in 2015 by daily newspapers through the channel of RSS flows explicitly titled "international" or "world". International RSS flows should have a universal coverage; that is, they would be likely to speak about all countries of the world without exception. They should also produce a regular number of foreign news stories per week over the year 2015

 We obtained a final list of 32 newspapers in three languages (16 english, 8 french;, 8 spanish) covering foreign news during the 52 weeks of the period of observation starting Monday, January 5, 2015, and ending Sunday, January 3, 2016 (Table 2).

---------------------------

 Add Table 2 here

 ----------------------------

This sample of media does not intend to be representative of the total volume of news circulating during the period of observation. It was based on the choice to maximize the diversity of geographical locations, under the constraint of the availability of newspapers fulfilling the conditions of comparability. One of the French media (El Watan, Algeria) was excluded because of too strong heterogeneity of news flow, reducing the sample to 31 newspapers (*See Supplementary material*).

## The 3-D cube and weighting options

From the 321269 news items, we derive a cube F(m,t,p) where each news story produced by media outlet (m) during week (t) is allocated to country (p). When several countries are present in the same item (e.g., "Putin discusses the Syrian crisis with Obama"), we divide the weight of the news between each of the countries mentioned at least once (e.g., 1/3 Russia, 1/3 USA, 1/3 Syria).

From the first cube, we derive a weighted cube Fw(m,t,p), where the number of news stories is inversely weighted by media and time to ensure an equal number of news stories sent by each media outlet during each week. This weighting scheme prevents the under-representation of "small" newspapers and gives the same weight in the analysis to the *Guardian* (55573 news) and to the *Chronicle of Zimbabwe* (2095 news). It also introduces a correction of the time sample, with an equal weight for all weeks.

## The salience of countries in the sample

Computing the number of news stories received by each country in 2015 (Table 3) demonstrates that the hierarchy of countries is not strongly modified by the weighting scheme, except for specific cases such as Australia, which represented 2.0% of news with raw data (ranked 9th) but only 0.9% of news (ranked 32th) after the weighting procedure. One must immediately understand that, weighted or not, the sample of news is not representative of world news and is clearly focused on Western media.

---------------------------

  Add Table 3 here

  ------------------------

Looking at the spatial distribution of news, we can notice a clear concentration of news in Northern America, Western Europe and Middle East (Figure 1).

---------------------------

  Add Figure 1 here

  ------------------------

## Results

The originality of the proposal is that it introduces a disaggregation of the model by media and by week (1) to analyze the effects of exceptional events on the parameters of the model and (2) to compare the model parameters from one media outlet to another. The implication is that the parameters of our model can be estimated from three different perspectives:

1. *Global estimation*: one single set of parameters ($\alpha_1, \alpha_2, \alpha_3\ldots$);

2. *Time estimation*: one set of parameters for each time period ($\alpha_{1,t}, \alpha_{2,t}, \alpha_{2,t}\ldots$).

3.  *Media estimation*: one set of parameters for each media outlet ($\alpha_{1,m}, \alpha_{2,m}, \alpha_{2,m}\ldots$); and

This disaggregation of the model is generally not used by authors who try to estimate the gravity parameters with a classical ordinary least squares (OLS) linear regression after logarithmic transformation of the equation. Indeed, the disaggregated model presents the apparent disadvantage of introducing a majority of zero values, which are difficult to manage with this family of statistical models. However, this disadvantage disappears if we use a count model such as Poisson regression with maximum likelihood estimation (IDRE, 2017; Long & Freese, 2006). Naturally, the problem is the risk of overdispersion, which is clearly at stake in the distribution of news per country ($\mu$ = 0.81, $\sigma$. = 4.37). Therefore, we have tested three count models for the global model (weighted or not) and concluded based on the Akaike information criterion (AIC) (Table 4) that the optimal solution is to use a negative binomial model[1].

--------------------------

Add Table 4 here

--------------------------

## The global model

All of the variables introduced into the model appear significant with the expected sign (Table 5). Geographical size and demographic size logically have a positive impact on the number of news stories, as they increase the probability of the appearance of an event of interest inside a country's territory. However, the world is not flat, and the flow of foreign news also decreases with distance and increases with cultural proximity (a common language). Concerning elite countries, we can observe a certain media preference for the richest countries associated with an obvious privileging of the P5 of the UNSC and the Holy See. However, the situation is less clear for other members of the G20, which benefit only from a small increase in media coverage, all things being equal with the other parameter. Finally, we can observe the existence of an important effect of time dependence.

--------------------------

-----------------------

[1] We have also tested a zero-inflated binomial regression model that produces better results for the global estimation but that was not adapted for estimation by media or time period.

Add Table 5 here

--------------------------

## The model segmented by week

What happens when the same parameter is calculated at the week level? Do we observe a stability of the parameters or of the significance? What is the effect of exceptional events such as the earthquake in Nepal or the Paris attacks from Charlie Hebdo and Bataclan? To answer these questions, we have computed the same negative binomial model by week and tested the significance of each parameter for each time period. Here, we analyzed the z-values to visualize the variation of the significance of the parameters (Figure 2).

----------------------------

Add Figure 2 here

----------------------------

### Size effects

Geographical size and population density display very similar profiles and are systematically associated with very strong positive effects during all weeks of the period. However, a remarkable decrease in the z-value can be observed in both curves during the week of March 9-15. Examining the data makes it is easy to demonstrate that this exception was associated with the severe Tropical Cyclone Pam from March 6 to 22. This storm was the second most intense tropical cyclone of the south Pacific Ocean in terms of sustained winds and is regarded as one of the worst natural disasters in the history of Vanuatu. However, it also affected other small island states, notably the Solomon Islands and Tuvalu.

In general, GDP per capita is also positively significant, but there are more exceptions, and we can observe several weeks where the economic development of countries is not associated with more news, all things being equal with other factors. Moreover, we observe that the effect is significantly negative during the two weeks corresponding to the initial shock and the first major aftershock of the Nepal earthquake.

The rule postulated by Galtung and Ruge about elite countries is therefore statistically true at the year level but not systematically true at the week level. Regardless, these two examples of Tropical Cyclone Pam and the Nepal earthquake well support the hypothesis formulated by Galtung and Ruge by crossing rules 9 and 12: "The lower the rank of the nation, the more negative will the news from that nation have to be" (Galtung & Ruge, 1965, p.83).

## Elite nation effects

As expected, the exceptional situation of the Holy See in foreign news is confirmed during all weeks of the year, which demonstrates a perfect example of a small state power strategy (Chong, 2010). The same is true for the P5 of the UNSC, but with more exceptions and even three weeks where the effect became negative. To some extent, we can argue that the year 2015 has artificially reinforced this effect, particularly because of the two Paris attacks on Charlie Hebdo (January 2015) and Bataclan (November 2015), which twice placed France at the top of the international news. However, we can ask whether it is artificial only as long as we can hardly find a year where none of the P5 was involved in a major event. Other countries suffered terrorist attacks in 2015 (e.g., Tunisia, Nigeria), but the effect in terms of media coverage was clearly less important. We must also bear in mind that the majority of major international news agencies (AP, Agence France-Presse, Reuters, Xinhua, ITAR-TASS) originate precisely from the P5 of the UNSC and are subject to a national bias effect in favor of home interests (Horvit, 2006).

## Proximity effects

The distance effect appears to be always positive and significant at the week level, confirming that, contrary to the hypothesis of Friedman (Friedman, 2005), the world is not flat and that we are still very far from the end of geography postulated by O'Brien (O'Brien, 1992). As noted in the discussion of hypotheses, this persistent effect of distance is not related to the omission of other variables such as a common border or a former colonial relation that became non-significant when combined with distance and language. However, it is certainly related to the voluntary exclusion of flow-based data (trade, migration, tourism) from our model. We maintain that explaining flows by flows introduces a serious issue into the model and introduces uncontrolled loops of causal effects.

The existence of a common language appears to have the second most relevant effect of proximity. It is always positive but not always significant, and therefore, it appears to be less systematic than simple geographical distance. The existence of this effect is clearly related to the introduction into our sample of media from hispanophone and francophone countries. It is not a pure effect of communication and is also associated with a colonial heritage and other forms of historical and cultural proximity. However, this effect also reveals the existence of a separate arena of debate in international world news. The maximum peak of common language effects corresponds to a major diplomatic crisis involving the USA at the beginning of March 2015 that was covered intensively by newspapers from Latin America and the USA but largely ignored elsewhere.

## The model segmented by media and "culture-bound factors"

Hitherto, we have assumed the existence of single parameters of salience and proximity effects for all media of our sample. However, we suspect that this existence is not the case, and therefore, it is important to compute separate models for each of our 31 newspapers to verify another assumption of Galtung and Ruge (1965) regarding the fact that some of the laws of foreign news are not necessarily identical in all media. More precisely, Galtung assumed that laws F1-F8 but not laws F9-F12 were culture-free:

*But there is little doubt that there are also culture-bound factors influencing the transition from events to news, and we shall mention four such factors that we deem to be important at least in the northwestern corner of the world.* (Galtung & Ruge, 1965, p.68)

The table of parameters obtained for each of the 31 newspapers (see the Appendix) reveals very important variations in the effect of variables that are not always significant and that can even display opposite signs compared to the general model. To briefly summarize this finding, we first conducted a principal component analysis to propose a synthetic visualization of the correlation between variables and the proximity between newspapers (Figure 3).

----------------------------

Add Figure 3 here

----------------------------

Hierarchical clustering using Ward's criteria was also performed in order to minimize intra-group and maximize inter-group variance of parameters (Rand, 1971). The tree of classification clearly reveals the existence of 4 types of media in terms of factors that influence the reporting of foreign news (Figure 4).

------------------------------

Add here Figure 4 or just below Figure 3 on a separated page

------------------------------

- **Type 1: A focus on elite countries and the British Commonwealth** is characteristic of six anglophone newspapers from the UK, Australia, India and Malta. The effect of the Vatican is strongly reduced, and geographical size and demographic size play a minor role in the choice to publish news stories compared to economic power. This profile is clearly characterized by a strong bias in favor of news concerning the USA and nations formerly belonging to the Empire

- **Type 2: A focus on elite countries and the Hispanic world** is characteristic of four newspapers from Chile, Mexico and Spain. Contrary to other hispanophone newspapers, they are not characterized by a preference for small and neighboring countries and for the Vatican. They produce a global coverage but with a bias in favor of countries with the same language.

- **Type 3: A focus on the largest emerging countries** is typical of anglophone newspapers located in North America as well as two Australian newspapers, China Daily and the Financial Times. This group of media is characterized by a specific interest in China and Russia and, more generally in the emerging powers of the 21st century. They are barely influenced by common language parameters, in contrast to the newspapers of type 1.

- **Type 4: A focus on neighboring countries** is typical of the majority of francophone and hispanophone newspapers as well as some anglophone newspapers located in small or poor countries. The newspapers of this group appear to be very influenced by proximity but are less focused on the richest countries or the P5 of the UNSC. They therefore offer a loop on parts of the world that are less covered by media of the three previous types.

## The model as filter for event detection through space and time

The exclusion of non-structural factors from our model present the great interest to *filter* the exceptional events occurring in space and time that has finally become news. We suggest two complementary approach for the discovery of such events.

### Identification of exceptional weeks

Based on the model disaggregated by time period, it is possible to analyze the variations of the explanatory power of the structural parameters, measured by the proportion of deviance explained. A brutal modification of the parameter of the model associated with an increase of decline in the quality of adjustment is the obvious signal that something exceptional is happening somewhere in the world as we can notice in Figure 5.

------------------------------

Add here (new) Figure 5

------------------------------

The results suggest an interesting hypothesis about the fact that exceptional events occurring in elite countries will reinforce the model (e.g. Paris attacks in January and November, Iran nuclear deal of July, UK storms of December, …). On the contrary exceptional events occurring in peripheral countries will reduce the explanatory power of the structural parameters (e.g. Bardo and Sousse attacks in March and June, Pam cyclone, Chile and Nepal earthquakes, …).

## Identification of exceptional countries

Based on the model disaggregated by media, it is possible to analyze the countries that have received an exceptional media attention. The residuals that are common to the majority of newspapers, despite their difference rules, make theoretically possible to identify places where major events occurred during the period of observation. In the case of the year 2015 we can easily identify on the events responsible from exceeding number of news in countries like Syria, Greece, Nepal, Yemen, Ukraine, Cuba or Tunisia (Figure 6)

-----------------------------

Add here (new) Figure 6

-----------------------------

It is therefore very clear that the introduction of the supplementary variables concerning events (Segev, 2016) would certainly has dramatically improved the explanatory power of the model, and we have made available on the website of the journal all the data and program used in the present paper for people interested by such development.

## Analysis the coverage of political crisis through space, time and media

The structural model proposed in these papers offers an innovative solution for the analysis of the media coverage of political crisis and the benchmarking of the international agenda of newspapers located in different countries (Neumann & Fahmy, 2016). A nice example is given by the civil war in Yemen that started in January 2015 when Houthi fighters seized the presidential compound in Sana'a. These initial event was followed by many developments, in particular the battle for the control of the Aden's airport in March 2015 that was followed by a Saudi Arabian-led intervention named "Decisive Storm" including many countries. It appears clearly from the residuals of our model that the interest of the majority of newspapers of our sample for the civil war in Yemen quickly declined after the Nepal's earthquake. Only a minority of newspapers maintained a regular focus on the conflict that turned to be a forgotten war in the majority of the media.

------------------------------

Add here (new) Figure 7

---------------------------

## Discussion

The main added value of our proposal has to be found in the attempt to propose an inductive approach of the discovery of international news event through a structural model where the gravity model proposed by Zipf (1946) and further developed by other authors (Tobler, 1995) is used as a filtering procedure. These approach is in our opinion complementary to the deductive approach developed by the majority of the specialist of International News Flow Theory (Kim & Barnett, 1996; Wu, 1998, 2000, 2003, 2007 ;  Shoemaker, 2006 ; Segev, 2010,2014,2017). The model that we propose is certainly less efficient in terms of explanatory power (because of the exclusion of event-based and flow-related factors) but present the advantage to be possibly applied in real time for the detection of emerging events. The simultaneous disaggregation of the model parameters by week, countries and media appears as an important innovation for the analysis of competition between news at international level (Grasland & al., 2016). It offers promising solution to the answer of many questions raised by specialist of agenda setting theory and competition between news (Östgaard, 1965 ; Peterson, 1981 ; Shoemaker & Reese, 2013, McCombs, 2018)

We have demonstrated that, despite the voluntary exclusion of classical explanatory variables (flows and conflicts), the classical rules of spatial interaction models used for modeling trade or migration can successfully be applied to the circulation of international news. To a large extent, it is possible to predict the salience of foreign countries in the international agenda of daily newspapers with a very simple sets of structural parameters consisting of the size (area, population, GDP), the remoteness (geographical distance, border, common language) and the elite status of countries (members of PM5, Holly See). This model also generally remains valid when it is disaggregated by short periods of time. Despite the perturbation induced by exceptional events such as the Paris attacks or the earthquake in Nepal, a large part of the international agenda of newspapers remains more likely to be predictable at the week level using the same parameters than during a one-year period. Indeed, we have demonstrated that these exceptional events can introduce a temporary modification of the values and significate of the parameters of gravity model. However, it is always the case that more than half of the production of international news remains subject to the rules of "business as usual".

However, it is important to consider the disaggregation of the model not only by weeks but also by media. Indeed, we have demonstrated that the the rules governing media exhibit spatial non-stationarity. These does not mean that we should adopt an idiographic approach considering that all media are unique. But we have to consider the interest of a place-based approach where universal rules certainly exist but display variation in their parameters (Goodchild & Janelle, 2004, p.8). Media located in elite countries – particularly in the USA – are relatively less influenced by geographical distance or a common language than are media located in poor countries. Elite media are therefore likely to produce a broader coverage of world news but with a higher level of conformism and reproduction than media located in more peripheral areas. Because of their focus on more local news, peripheral newspapers can zoom in some parts of the world that are less covered by mainstream media. It is therefore of particular interest to carefully examine the residuals produced by each media outlet facing an event and to avoid a blind aggregation of news produced by different media. Finally, the most important interest of our approach is to offer a disaggregated vision of the variations of salience of countries by media and through time, making possible to reveal strong divergences in international agenda according to language and regional localization of newspapers.

In a previous research project called GEOMEDIA, we have developed a tool called "event explorer" for the interactive application for the visualization of international event through space and time, presented at the final conference of the project (Lambert & al., 2016) and freely available on the web[2]. Based on the same data than the present paper, this tool focus on the detection of events in the narrow sense of brutal increase of the interest of a media for a country as compared to previous time period. The tool was very useful for the identification of "new" event creating a sudden break in time series, as we can see with the example of the first week of January 2015 where all media reported the Charlie Hebdo attack in Paris (Figure 7). But the tool was not able to identify structural anomalies associated to long term crisis over several months or years, like it was the case for Syria, Greece or Ukraine during the same period.

The residuals of the model presented here proposes a solution to the problem because they are based on structural parameters that are stable through time. The model can therefore identify countries that are characterized by a permanent media coverage (like the US or the Australia in the majority of the media of our sample) or media that received a coverage more important than usual during several months (Yemen, Ukraine) or years (Syria, Greece) because a situation of long-term political crisis. But the model can also indicate what are the variation of the pattern of time coverage of countries from one media to another.

As this paper was conceived as a tribute to the seminal publications of Galtung & Ruge (1965), we cannot resist the temptation to present finally our main conclusion in the form of an update of Galtung's famous metaphor of the world as giant broadcast station:

*Imagine that the network of world media can be likened to an enormous set of event sensors, with each tracking signals emitted from all countries of the world but with a specific wavelength. The reception is continuous, corresponding to the truism that something is always happening in the world. Even if each media outlet consider itself to be "objective", what it chooses to consider an "event" is culturally determined. The emerging salience of world events, then, is strongly variable, depending on the results that each media outlet obtains by scanning the dial with one's cultural bias, and it is particularly confusing when it concerns events occurring in small and medium-sized countries. Obviously, the quantitative aggregation of this cacophony does not make sense; it may become meaningful only if one sensor is compared to others and listened to for some period of time to identify normal behavior and exceptions. Since we cannot register all media of the world, we have to select them to be as different as possible in terms of cultural bias, and the question is what are the few exceptional events that strike the attention of all of them.*

---

[2] http://geomediaex.ums-riate.fr/

## Acknowledgement

This project has received funding from the European Union's Horizon 2020 research and innovation program under grant agreement no. 732942. He has also benefit from the support of the French research plateform Huma-Num (https://www.huma-num.fr/) for the collection of storage of millions of news extracted from RSS flows from 2014 to present.

We would also like to thank both reviewers for their insightful comments on the paper, as these comments led us to an improvement of the work, especially concerning the state of the art and the enlargement of the initial perspective in the discussion of the results.

# Tables

Table 1: A reminder of the 12 laws governing the structure of foreign news postulated by Galtung & Ruge (1965)

| Factor | Hypothesis |
| --- | --- |
| FI | If the frequency of the signal is outside the dial it will not be recorded. |
| F2 | The stronger the signal, the greater the amplitude, the more probable that it will be recorded as worth listening to |
| F3 | The more clear and unambiguous the signal (the less noise there is), the more probable that it will be recorded as worth listening to |
| F4 | The more meaningful the signal, the more probable that it will be recorded as worth listening to |
| F5 | The more consonant the signal is with the mental image of what one expects to find, the more probable that it will be recorded as worth listening to |
| F6 | The more unexpected the signal, the more probable that it will be recorded as worth listening to |
| F7 | If one signal has been tuned in to the more likely it will continue to be tuned in to as worth listening to |
| F8 | The more a signal has been tuned in to, the more probable that a very different kind of signal will be recorded as worth listening to next time |
| F9 | The more the event concerns elite nations, the more probable that it will become a news item |
| F10 | The more the event concerns elite people, the more probable that it will become a news item |
| F11 | The more the event can be seen in personal terms, as due to the action of specific individuals, the more probable that it will become a news item |
| F12 | The more negative the event in its consequences, the more probable that it will become a news item |

Source : Galtung, J., & Ruge, M. H. (1965). The structure of foreign news the presentation of the Congo, Cuba and Cyprus Crises in four Norwegian newspapers. *Journal of Peace Research*, *2*(1), 64–90.

Table 2: Definition of the sample of 31 daily newspapers under studies

| Code | Name | lang. | Country | URL | Nb. Items |
|------|------|-------|---------|-----|-----------|
| en_AUS_austr | The Australian | en | AUS | www.theaustralian.com.au | 5945 |
| en_AUS_dtele | The Daily Telegraph (AUS) | en | AUS | www.dailytelegraph.com.au/ | 5881 |
| en_AUS_moher | The Sydney Morning Herald | en | AUS | www.smh.com.au/ | 16916 |
| en_AUS_theag | The Age | en | AUS | www.theage.com.au/ | 16675 |
| en_CAN_starc | The Star | en | CAN | www.thestar.com | 7292 |
| en_CAN_vansu | The Vancouver Sun | en | CAN | www.vancouversun.com | 5096 |
| en_CHN_china | China Daily | en | CHN | www.chinadaily.com.cn | 11730 |
| en_GBR_daily | The Daily Telegraph (UK) | en | GBR | www.telegraph.co.uk/ | 19054 |
| en_GBR_finat | The Financial Times | en | GBR | www.ft.com | 9073 |
| en_GBR_guard | The Guardian | en | GBR | www.theguardian.com/ | 55573 |
| en_IND_tindi | The Times of India | en | IND | timesofindia.indiatimes.com | 12784 |
| en_MLT_tmalt | The Times of Malta | en | MLT | www.timesofmalta.com | 7115 |
| en_USA_latim | The Los Angeles Times | en | USA | www.latimes.com/ | 5362 |
| en_USA_nytim | The New York Times | en | USA | www.nytimes.com | 14963 |
| en_USA_usatd | USA Today | en | USA | www.usatoday.com/ | 8535 |
| en_ZWE_chron | The Chronicle | en | ZWE | www.chronicle.co.zw | 2095 |
| es_BOL_patri | La Patria | es | BOL | lapatriaenlinea.com | 3410 |
| es_CHL_terce | La Tercera | es | CHL | www.latercera.com/ | 7357 |
| es_ESP_catal | El Periodico de Catalunya | es | ESP | www.elperiodico.com/es/ | 7594 |
| es_ESP_elpai | El Pais | es | ESP | elpais.com/ | 13699 |
| es_ESP_farod | Faro de Vigo | es | ESP | www.farodevigo.es/ | 2971 |
| es_MEX_croni | La Cronica de Hoy | es | MEX | www.cronica.com.mx/ | 6195 |
| es_MEX_Infor | El Informador | es | MEX | www.informador.com.mx | 10802 |
| es_VEN_unive | El Universal (VEN) | es | VEN | www.eluniversal.com/ | 17663 |
| fr_BEL_derhe | Derniere Heure | fr | BEL | www.dhnet.be/ | 4367 |
| fr_BEL_lesoi | Le Soir | fr | BEL | www.lesoir.be | 5850 |
| fr_FRA_antil | France Antilles | fr | FRA | www.franceantilles.fr | 11017 |
| fr_FRA_figar | Le Figaro | fr | FRA | www.lefigaro.fr/ | 5344 |
| fr_FRA_lepar | Le Parisien | fr | FRA | www.leparisien.fr/ | 5061 |
| fr_FRA_liber | Liberation | fr | FRA | www.liberation.fr/ | 5299 |
| fr_FRA_lmond | Le Monde | fr | FRA | lemonde.fr | 10551 |

Source : Data on RSS flows collected by research project ANR Geomedia, with the support of TGIR Huma-Num. Statistical analysis realized by author with statistical software R.

Table 3: Salience of countries in international news of 31 daily newspapers in 2015

| | Country | Raw data | | | Standardized data | | |
|---|---|---|---|---|---|---|---|
| ISO3 | Name | nb | freq | rank | nb | freq | rank |
| USA | United States of America | 38878 | 15,97 | 1 | 37042 | 15,19 | 1 |
| FRA | France | 15106 | 6,20 | 2 | 14908 | 6,11 | 2 |
| GBR | United Kingdom | 8714 | 3,58 | 6 | 10371 | 4,25 | 3 |
| SYR | Syria | 9098 | 3,74 | 4 | 9521 | 3,91 | 4 |
| GRC | Greece | 9698 | 3,98 | 3 | 9467 | 3,88 | 5 |
| RUS | Russian Federation | 8656 | 3,55 | 7 | 9077 | 3,72 | 6 |
| CHN | China | 9075 | 3,73 | 5 | 8570 | 3,52 | 7 |
| DEU | Germany | 6760 | 2,78 | 8 | 6799 | 2,79 | 8 |
| TUR | Turkey | 4744 | 1,95 | 10 | 5127 | 2,10 | 9 |
| ISR | Israel | 4651 | 1,91 | 11 | 4973 | 2,04 | 10 |
| IRQ | Iraq | 4221 | 1,73 | 12 | 4693 | 1,92 | 11 |
| IRN | Iran | 4211 | 1,73 | 13 | 4044 | 1,66 | 12 |
| JPN | Japan | 3649 | 1,50 | 15 | 3813 | 1,56 | 13 |
| VAT | Holy See | 3538 | 1,45 | 17 | 3716 | 1,52 | 14 |
| EGY | Egypt | 3403 | 1,40 | 19 | 3605 | 1,48 | 15 |
| UKR | Ukraine | 3383 | 1,39 | 20 | 3579 | 1,47 | 16 |
| ESP | Spain | 3926 | 1,61 | 14 | 3568 | 1,46 | 17 |
| MEX | Mexico | 3563 | 1,46 | 16 | 3526 | 1,45 | 18 |
| ITA | Italy | 3495 | 1,44 | 18 | 3490 | 1,43 | 19 |
| AFG | Afghanistan | 3139 | 1,29 | 22 | 3231 | 1,33 | 20 |
| ARG | Argentina | 2983 | 1,22 | 24 | 3147 | 1,29 | 21 |
| IND | India | 3199 | 1,31 | 21 | 3100 | 1,27 | 22 |
| NPL | Nepal | 2813 | 1,16 | 25 | 2937 | 1,20 | 23 |
| NGA | Nigeria | 2290 | 0,94 | 31 | 2910 | 1,19 | 24 |
| BRA | Brazil | 3028 | 1,24 | 23 | 2863 | 1,17 | 25 |
| IDN | Indonesia | 2731 | 1,12 | 26 | 2739 | 1,12 | 26 |
| YEM | Yemen | 2548 | 1,05 | 28 | 2705 | 1,11 | 27 |
| PSE | Palestine | 2577 | 1,06 | 27 | 2636 | 1,08 | 28 |
| BEL | Belgium | 2445 | 1,00 | 29 | 2437 | 1,00 | 29 |
| VEN | Venezuela | 1935 | 0,79 | 35 | 2418 | 0,99 | 30 |
| SAU | Saudi Arabia | 2410 | 0,99 | 30 | 2416 | 0,99 | 31 |
| AUS | Australia | 4871 | 2,00 | 9 | 2298 | 0,94 | 32 |
| COL | Colombia | 2257 | 0,93 | 33 | 2290 | 0,94 | 33 |
| CUB | Cuba | 2278 | 0,94 | 32 | 2272 | 0,93 | 34 |
| TUN | Tunisia | 1858 | 0,76 | 37 | 2124 | 0,87 | 35 |
| ZAF | South Africa | 1738 | 0,71 | 39 | 2102 | 0,86 | 36 |
| LBY | Libya | 1876 | 0,77 | 36 | 2091 | 0,86 | 37 |
| PAK | Pakistan | 2241 | 0,92 | 34 | 2068 | 0,85 | 38 |
| CHL | Chile | 1605 | 0,66 | 40 | 1719 | 0,71 | 39 |
| KOR | Republic of Korea | 1466 | 0,60 | 41 | 1500 | 0,62 | 40 |

Source : Data on RSS flows collected by research project ANR Geomedia, with the support of TGIR Huma-Num. Statistical analysis realized by author with statistical software R.

Table 4: Choice of the best statistical model

| Count model | Data | df | AIC |
|---|---|---|---|
| Poisson | Raw | 11 | 622430,2 |
| Poisson | Normalized | 10 | 700908,6 |
| Negative Binomial | Raw | 12 | 427047,8 |
| Negative Binomial | Normalized | 11 | 450361,6 |
| Zero Inflated Poisson | Raw | 22 | 531749,0 |
| Zero Inflated Poisson | Normalized | 20 | 525232,8 |

Source : Data on RSS flows collected by research project ANR Geomedia, with the support of TGIR Huma-Num. Statistical analysis realized by author with statistical software R.

Table 5: Parameters of global model

| Parameter | Estimate | Std. Error | z value | Pr(>|z|) |
|---|---|---|---|---|
| (Intercept) | -3,735 | 0,076 | -49,467 | <0.0001 |
| V1_sup | 0,505 | 0,004 | 118,924 | <0.0001 |
| V2_dem | 0,553 | 0,005 | 104,448 | <0.0001 |
| V3_eco | 0,184 | 0,005 | 35,909 | <0.0001 |
| V4_PM5 | 0,633 | 0,029 | 22,049 | <0.0001 |
| V5_G14 | 0,077 | 0,019 | 3,983 | <0.0001 |
| V6_VAT | 5,163 | 0,066 | 77,669 | <0.0001 |
| V7_dis | 0,333 | 0,007 | 49,527 | <0.0001 |
| V8_lan | 0,331 | 0,012 | 27,045 | <0.0001 |
| V9_time | 1,818 | 0,012 | 149,737 | <0.0001 |

Source : Data on RSS flows collected by research project ANR Geomedia, with the support of TGIR Huma-Num. Statistical analysis realized by author with statistical software R.

# Figures

Figure 1: Spatial distribution of international news in RSS flows of 31 daily newspapers in 2015

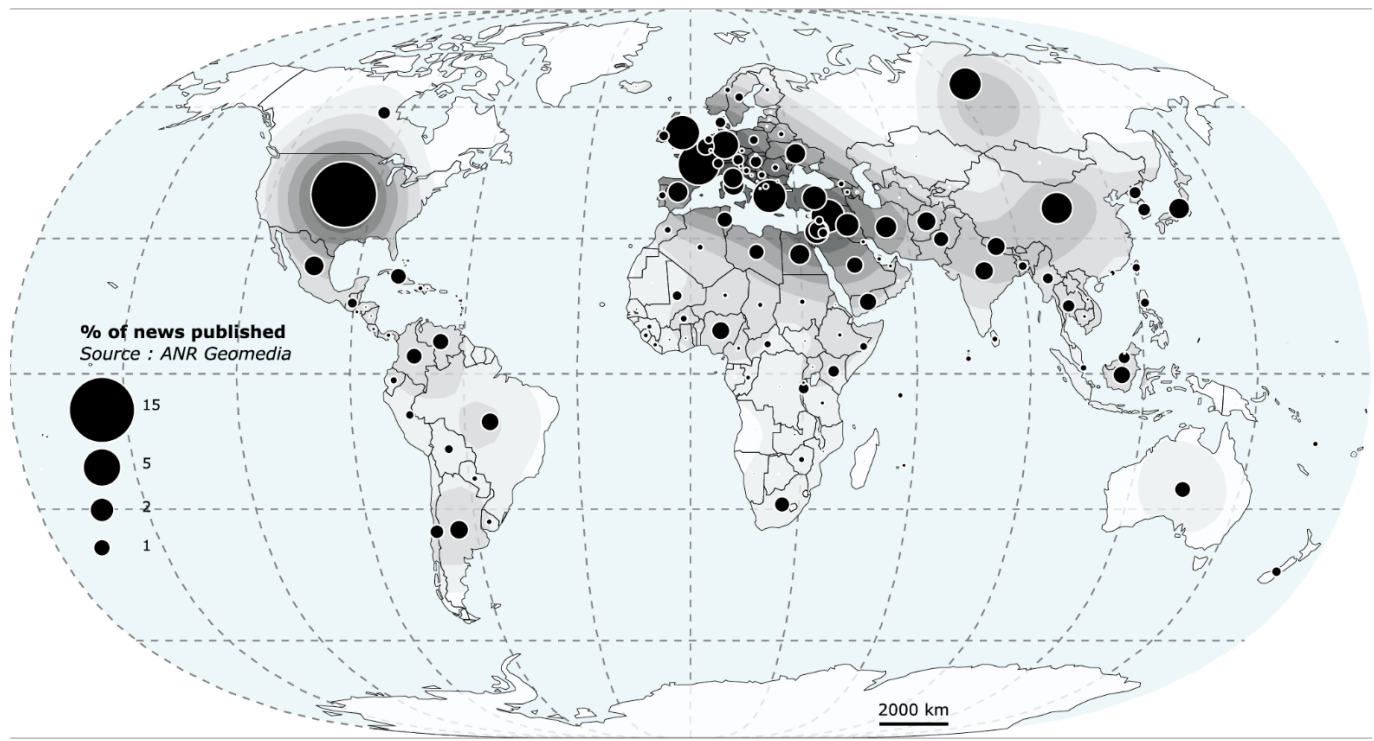

Source : Data on RSS flows collected by research  project ANR Geomedia, with the support of TGIR Huma-Num. Map realized with the interactive software Magrit (http://magrit.cnrs.fr/). High resolution figure :  *icg2018_4_fig_1.pdf*

Figure 2: Time variation of the determinant of international news flows during the 52 weeks of the year 2015

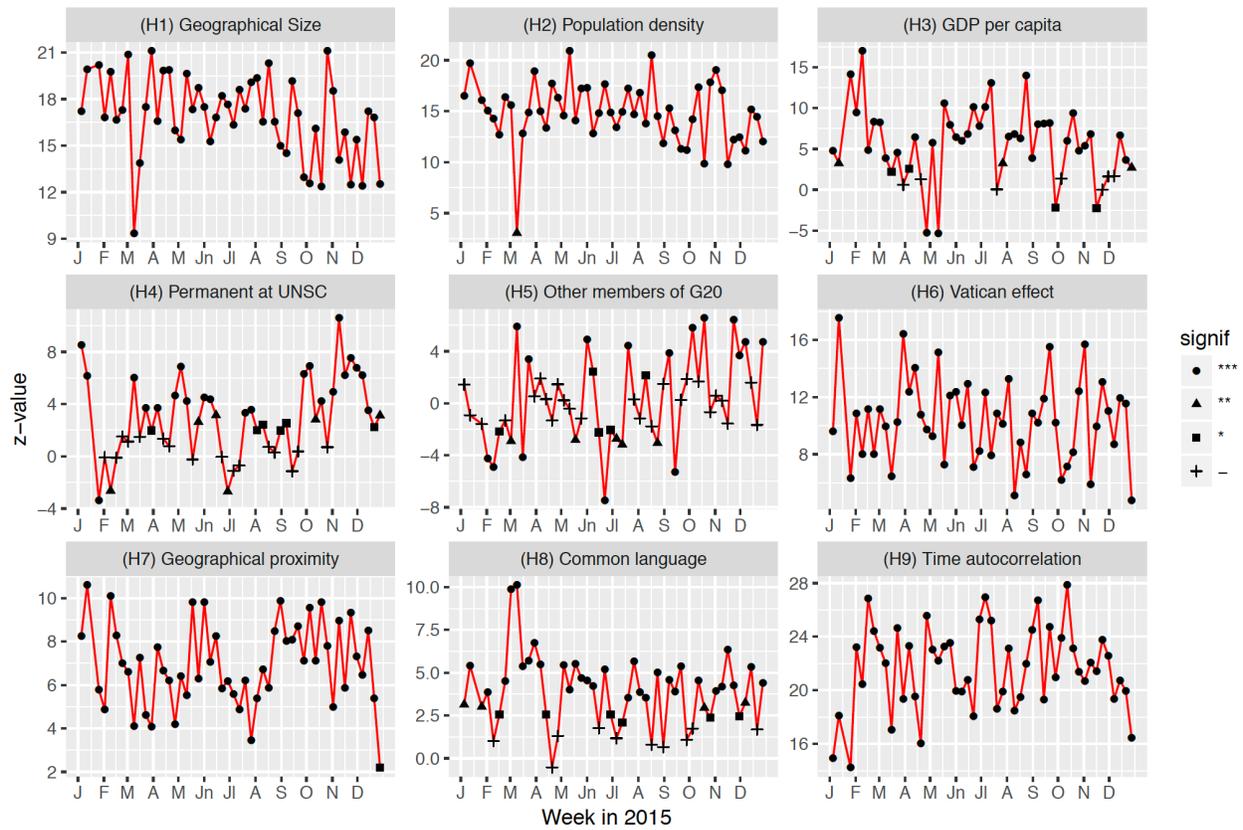

Source : Data on RSS flows collected by research project ANR Geomedia, with the support of TGIR Huma-Num. Statistical analysis realized by author with statistical software R.
High resolution figure : *icg2018_4_fig_2.pdf*

Figure 3: PCA on the determinant of international news flow for 31RSS flows of daily newspaper in 2015

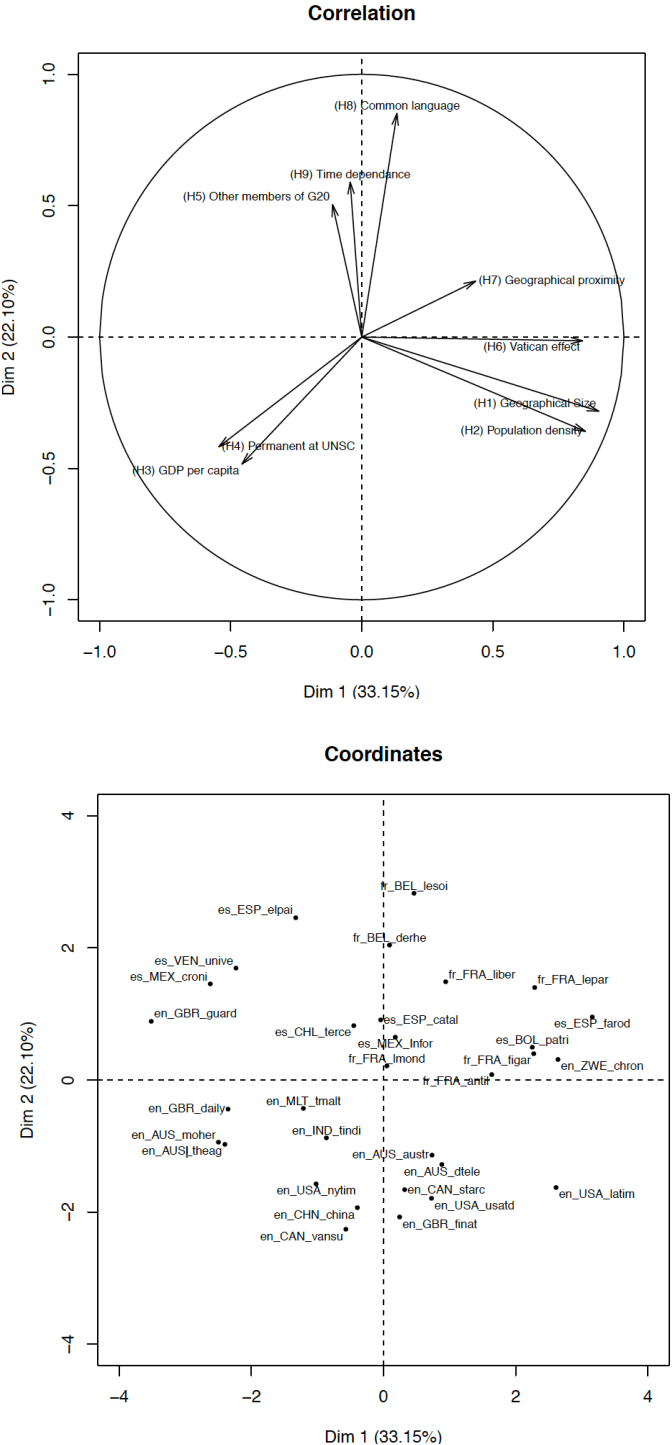

Source : Data on RSS flows collected by research project ANR Geomedia, with the support of TGIR Huma-Num. Statistical analysis realized by author with statistical software R.

High resolution figure : *icg2018_4_fig_3.pdf*

Figure 4: Cluster Analysis on the determinant of international news flow for 31RSS flows of daily newspaper in 2015

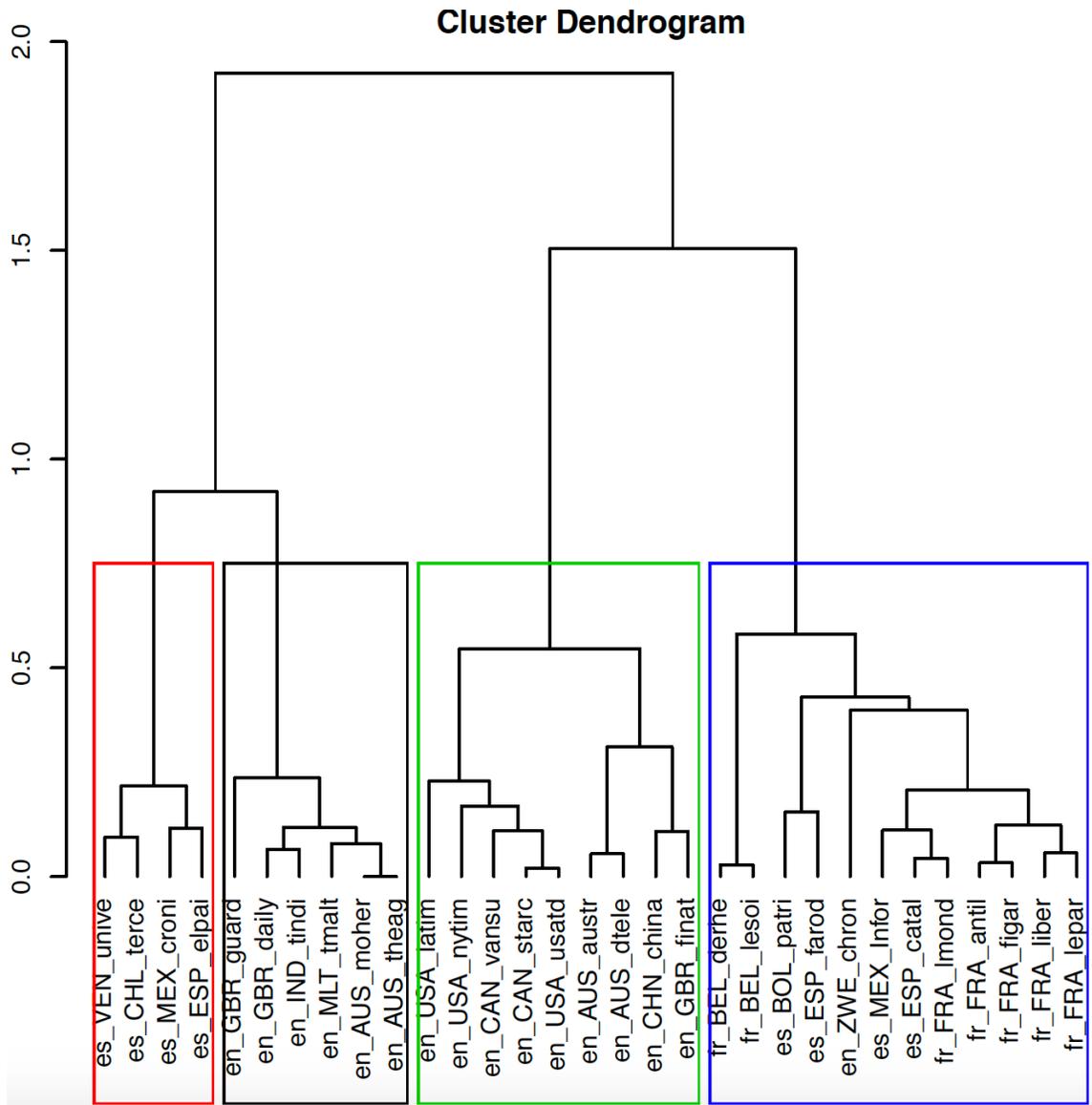

Source : Data on RSS flows collected by research project ANR Geomedia, with the support of TGIR Huma-Num. Statistical analysis realized by author with statistical software R.
High resolution figure : *icg2018_4_fig_4.pdf*

Figure 5: Explanatory power of the model during the 52 weeks of the year 2015

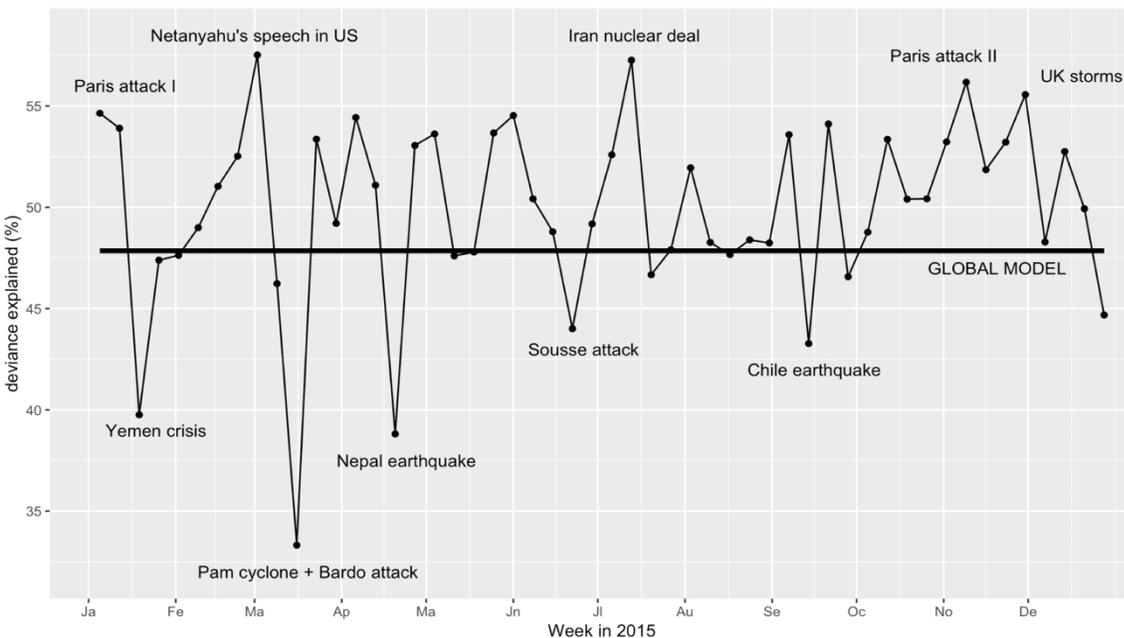

Source : Data on RSS flows collected by research  project ANR Geomedia, with the support of
TGIR Huma-Num. Statistical analysis realized by author with statistical software R.
High resolution figure :  *icg2018_4_fig_5.pdf*

Figure 6: Residual salience of world countries according to international RSS flows of 31 daily newspapers in 2015

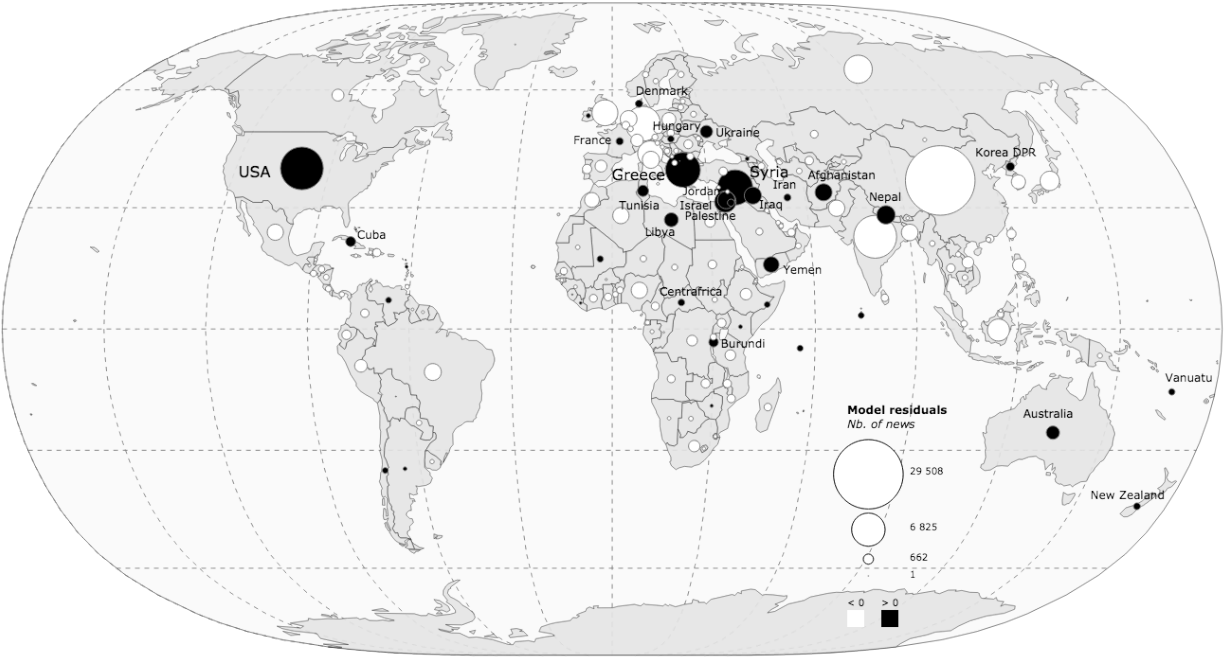

<u>Comment</u> : the map represent the absolute difference between the observed number of news received by a country and the predicted number of news according to a gravity like-model disaggregated by media and weeks.

Source : Data on RSS flows collected by research  project ANR Geomedia, with the support of TGIR Huma-Num. Statistical analysis realized by author with statistical software R.
High resolution figure :  *icg2018_4_fig_6.pdf*

Figure 7: Media coverage of Yemen by international RSS flows of 31 daily newspapers in 2015

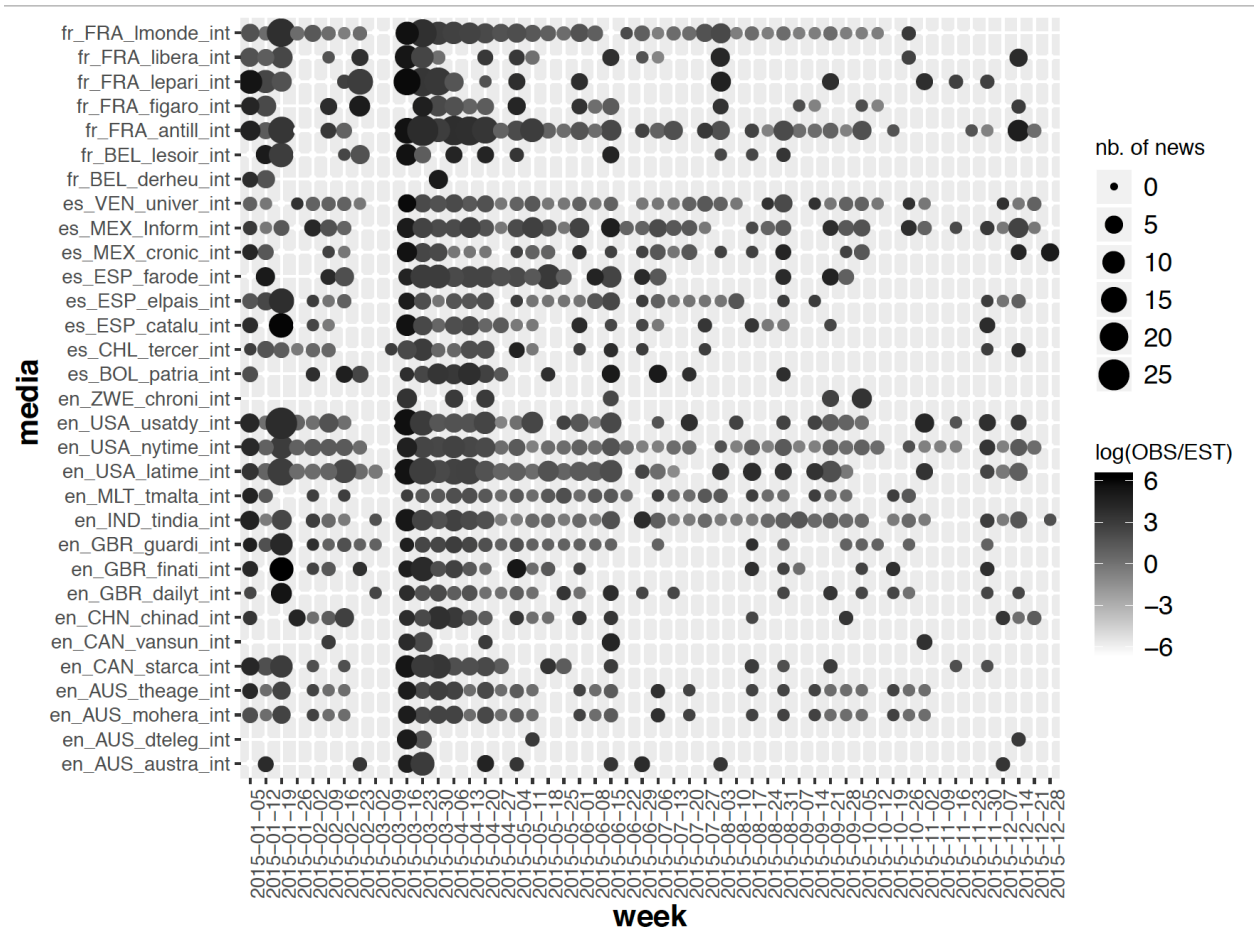

<u>Comment</u>: the size of dots represents the number of news published by each media about Yemen for each of the weeks of the year 2015. The level of gray represents the intensity of news according to a gravity-like model

Source : Data on RSS flows collected by research project ANR Geomedia, with the support of TGIR Huma-Num. Statistical analysis realized by author with statistical software R.
High resolution figure : *icg2018_4_fig_7.pdf*